\documentclass[journal=jctcce,manuscript=article]{achemso}

\usepackage[version=3]{mhchem} 

\usepackage{bm}
\usepackage{graphicx}

\newcommand{\subscript}[1]{\ensuremath{_{\textrm{\footnotesize{#1}}}}}

\author{J. P. Coe}
\email{J.Coe@hw.ac.uk}
\affiliation{Institute of Chemical Sciences, School of Engineering and Physical Sciences, Heriot-Watt University, Edinburgh, EH14 4AS, United Kingdom}

\title[\texttt{achemso} ]{Machine learning configuration interaction for {\it ab initio} potential energy curves}

\begin{document}
\begin{abstract}
The concept of machine learning configuration interaction (MLCI) [{\it J. Chem. Theory Comput.} {\bf 2018}, {\it 14}, 5739], where an artificial neural network (ANN) learns on the fly to select important configurations, is further developed so that accurate {\it ab initio} potential energy curves can be efficiently calculated. This development includes employing the artificial neural network also as a hash function for the efficient deletion of duplicates on the fly so that the singles and doubles space does not need to be stored and this barrier to scalability is removed. In addition configuration state functions are introduced into the approach so that pure spin states are guaranteed, and the transferability of data between geometries is exploited. This improved approach is demonstrated on potential energy curves for the nitrogen molecule, water, and carbon monoxide. The results are compared with full configuration interaction values, when available, and different transfer protocols are investigated. It is shown that, for all of the considered systems, accurate potential energy curves can now be efficiently computed with MLCI. For the potential curves of N\subscript{2} and CO, MLCI can achieve lower errors than stochastically selecting configurations while also using substantially less processor hours. 
\end{abstract}


\section{Introduction}

The use and success of machine learning is rapidly increasing in many fields, and quantum chemistry is no exception to this. In this context its impressive applications have often involved interpolating and extrapolating potential energy surfaces for dynamics calculations. This includes recent work using Gaussian process regression\cite{Qu18,Schmitz18,Wiens19} or artificial neural networks (ANNs).\cite{Gastegger15} 

Alternatively one may employ machine learning, often trained on density-functional theory (DFT) results, to predict properties from electronic structure data.
For example, using DFT data, an efficient deep learning approach has been developed and demonstrated on molecules up to the size of salicylic acid where it accurately predicts energies.\cite{Schutt17} ANNs have also been trained to predict spin-state splittings of transition metal complexes.\cite{Janet17} Various machine learning methods were investigated for the prediction of atomization energies from molecular geometries in Ref.~\citenum{Hansen13}. Then a new representation for the data was introduced\cite{Faber18} to enable machine learning based on 
kernel ridge regression to achieve chemical accuracy for organic compounds. A novel machine learning algorithm similar to an ANN was put forward in Ref.~\citenum{Gubaev18} together with an active learning approach that chooses new additions to the training set to reduce errors, which enabled chemical accuracy in prediction to be reached faster.

Machine learning, trained to predict properties, has also been used to create functionals for DFT. For example, a convolutional neural network was developed to predict the kinetic energy for hydrocarbons when using the electron density as an input.\cite{Yao16} Machine learning was harnessed in Ref.~\citenum{Brockherde17} to predict the electron density from the external potential and also the energy from the electron density. A kinetic energy density functional has also been created using a multilayer ANN where the density and its gradients are used as the input to give predictions that are more accurate than existing functionals.\cite{Seino18} While an approach has been developed\cite{Custodio19} that trains an ANN to accurately predict the energies of homogeneous
Hubbard models then uses this as a functional for lattice-DFT calculations.

The elegant wavefunction approaches of second-order M{\o}ller-Plesset perturbation theory (MP2)\cite{MP2} and coupled-cluster singles and doubles (CCSD)\cite{CCSD} have also been used to provide training data for machine learning. However, although efficient and accurate for
problems that are well-described by small corrections to a single determinant, MP2 and CCSD can exhibit pathological behavior\cite{RevModPhysBartlett07} when confronted with multireference situations such as stretched bonds. The $\Delta$-Machine learning approach, trained on large numbers of organic molecules, has been demonstrated to accurately predict MP2 and coupled-cluster correlation energies.\cite{Ramakrishnan15} Ref.~\citenum{McGibbon17} used an ANN trained on CCSD(T) data to improve MP2 results by predicting corrections to the energies.
The CCSD or MP2 amplitudes have also been used as inputs for machine learning to predict the CCSD correlation energy.\cite{Margraf18} Ref.~\citenum{Welborn18} developed a machine learning approach using Gaussian process regression to predict correlation energies for MP2 and CCSD when using properties of the Hartree-Fock molecular orbitals as input data. There the transferability of the approach was demonstrated across geometries and when varying the structure or elements of molecules.

When using wavefunction methods, rather than DFT, generating sufficient data for training can be challenging, particularly if more computationally expensive methods that can cope with multireference problems are used. A third way is then to embed a machine learning approach within an electronic structure method with the aim of accelerating the {\it ab initio} calculation and improving its accuracy. This was done in the approach of machine learning configuration interaction (MLCI)\cite{MLCI} where an ANN learns on the fly to choose important configurations in an iterative selected configuration interaction (CI) scheme. This was demonstrated to significantly reduce the number of iterations to convergence and, for higher-accuracy single-point calculations, required less time than other ways of selecting configurations. A machine learning approach that successfully accelerates CCSD calculations has also recently been developed.\cite{Townsend2019}  There the k-nearest neighbors algorithm is used to predict the initial amplitudes from electronic structure properties when using training data from different geometries. In addition, the concept
of using machine learning to predict adaptive basis sets has been put forward\cite{Schutt18} and demonstrated to substantially accelerate DFT calculations. In this work we build upon the approach of MLCI\cite{MLCI}  with the aim of accelerating convergence whilst maintaining accuracy across a range of geometries when calculating {\it ab initio} potential energy curves. Consistently describing the energy across geometries that range from single reference to strongly multireference problems is a much greater challenge than making a single-point energy calculation more efficient. The new developments to overcome this include using the ANN also as a hash function when removing duplicates to increase the scalability of MLCI, employing configuration state functions to ensure pure spin states, and exploiting the transferability of data between geometries. These improvements aim to allow this selected CI approach to efficiently and accurately calculate {\it ab initio} potential energy curves.

Selected CI approaches seek to iteratively construct a compact wavefunction that sufficiently captures the properties of the full configuration interaction (FCI) wavefunction. Although the FCI wavefunction is the most accurate for a given basis, it is computationally intractable
for all but the smallest systems and basis sets. Yet often the majority of configurations have negligible coefficients in the FCI wavefunction. It is this sparsity that selected CI seeks to exploit so that, in principle, it can describe both single reference and multireference problems
sufficiently well using only a very small fraction of the FCI space without recourse to choosing an active space. One early example of selected CI was the method of Configuration Interaction using a Perturbative Selection made Iteratively (CIPSI).\cite{CIPSI} 
Recently the development and application of selected CI approaches has become increasingly popular. For example, rather than the exact first-order perturbation calculation of the wavefunction, Heat bath CI\cite{Holmes16} uses an approximation to this to choose configurations and has been  successfully applied to the potential energy curve for C\subscript{2} and a geometry of the chromium dimer. The adaptive sampling CI approach\cite{Tubman16} solely considers configurations with large enough coefficients when creating the singles and doubles space from which configurations are then selected using perturbation theory. In the MCI3 method,\cite{Ohtsuka17} the first-order correction to the wavefunction is sampled using diffusion Monte Carlo in configuration space\cite{ProjectorMCNagase08,AlaviFCIQMC1} to select determinants. MCI3 has been successfully applied to excited potential curves of the carbon dimer.\cite{Ohtsuka17}. This approach has been further developed\cite{Ohtsuka2019} to enable various pure spin states to be calculated simultaneously then demonstrated on multiple states of C\subscript{2} and an iron-sulfur cluster.
In the Quantum Package 2.0 program\cite{QuantumPackage2.0} an efficient CIPSI is presented where the second-order perturbation of the energy is used to select configurations. This approach was employed for the calculation of benchmark vertical transition
energies for a range of small compounds.\cite{Loos18} It has also been used to construct the nodal surfaces for diffusion Monte Carlo to enable very accurate calculations that, for example, gave the correct ground spin state of FeS in Ref.~\citenum{Scemama18} and have provided excitation energies for formaldehyde and water.\cite{Scemama18JCP} This implementation of CIPSI has been further developed to efficiently create pure spin states\cite{Applencourt18} and to provide a dressed perturbation approach.\cite{Garniron18}

Alternative methods to using perturbation theory to select configurations have also been created. For example, the $\Lambda$-CI approach\cite{Evangelista14} uses the expectation value of the configuration's energy.
The adaptive configuration interaction approach\cite{Schriber16} then considers both a perturbative estimate of the energy and the configuration's coefficient in the resultant wavefunction. Further development of this method has enabled the efficient computation
of excitation energies.\cite{Schriber17} In Monte Carlo configuration interaction (MCCI)\cite{MCCIGreer95,mcciGreer98,mccicodeGreer} configurations are chosen randomly and those whose absolute coefficient in the resultant wavefunction is less than a cutoff are eventually removed from it.
This approach has been actively further developed to successfully model, for example, vertical excitation energies,\cite{GreerMcciSpectra} hyperpolarizabilities,\cite{MCCIhyper} X-ray emission and absorption,\cite{XrayMCCI15} molecular tunnel junctions,\cite{GreerComplexMCCI}  dissociation energies via perturbation,\cite{MCCIfirstrowdiss} and spin-orbit splitting.\cite{Murphy18} Instead of using standard diagonalization, CDFCI\cite{Wang19} reformulates the FCI equation and uses coordinate descent where new determinants are only included if their update in the scheme is larger than a threshold. CDFCI is shown to compute highly accurate results for systems including water using $1.8\times 10^{6}$ determinants in the cc-pVDZ basis, and all-electron Cr\subscript{2}. Related to the concept of selected CI is the @CC method \cite{adaptiveCC2010} which uses truncated CI calculations to estimate the importance of orbitals and decide which to include in the coupled cluster calculations. While in the {\it a priori} identification of configurational deadwood approach\cite{Bytautas09} the form of the wavefunction is assumed to be similar to truncated coupled cluster then used to estimate the configurations to include based on truncated CI calculations. There are also the recently introduced approaches of incremental FCI\cite{Zimmerman17} and many-body expanded FCI\cite{Eriksen17,Eriksen18,Eriksen19} where the latter method allows the FCI energy be approached in a controlled and intrinsically parallel way by implementing multiple complete active space configuration interaction computations using a screening procedure as the numbers of orbitals are increased. For selected CI, if we knew which of the interacting configurations (single and double substitutions) were actually important then we could quickly build a compact and accurate wavefunction. This is exactly what the ANN in MLCI\cite{MLCI} attempts to learn to do as it orchestrates the iterative selected CI calculation.

In this Article, we briefly recap the approach of MLCI then present the use of the ANN as a hash function to delete duplicates on the fly and make the method more scalable by removing the requirement to store all of the single and double substitutions.
The adaptation of MLCI to use configuration state functions (CSFs) is then discussed followed by the three transfer protocols investigated in this Article. We first use Slater determinants and consider a single geometry of N\subscript{2} in the cc-pVDZ basis. Here MLCI when using the ANN as a hash function to remove duplicates is compared with sorting all of the single and double
substitutions, and also with stochastic selection (MCCI). We then use this more scalable MLCI with CSFs to calculate the potential energy curve of N\subscript{2} in the cc-pVDZ basis which we compare with FCI. Transferability is investigated for each of the three protocols and the accuracy is quantified. This and the time required is compared with previous MCCI results. The multireference character at selected geometries is also calculated. Next we turn to the potential energy curve for the double hydrogen dissociation of water in the cc-pVDZ basis and investigate the transfer protocols for MLCI when contrasted with FCI results. Finally we consider the potential energy curve for carbon monoxide in a cc-pVDZ basis and compare the FCI data, when available, with MLCI calculations using each of the transfer protocols.

\section{Methods}
\subsection{Machine learning configuration interaction}
The MLCI method\cite{MLCI} trains an ANN on the fly to predict important configurations in an iterated selected CI scheme. The inputs to the ANN are the occupations of the spin orbitals in the configuration of
interest which are either zero or one, together with one constant input. Hence for a basis set of $M$ orbitals there are $2M+1$ inputs. We use a single hidden layer of $n_{h}$ nodes, and one output which
gives the predicted transformed coefficient of the configuration. The value of the hidden nodes and output are given by logistic functions which produce values on $(0,1)$ and their weights are found when training on the results of each diagonalization by using stochastic gradient descent. 
This attempts to minimize the error $\frac{1}{2}(\text{output}-o_{t})^{2}$ where $o_{t}$ is the target transformed coefficient value for each configuration in the training set, which is shuffled after each complete pass through. Previously\cite{MLCI} the training in MLCI used 2000 passes through the data
then kept the weights that gave the lowest error on the verification set. As a MLCI calculation progresses the best weights often occur towards the start of these 2000 passes so now, for efficiency, the algorithm checks every 10 passes to see if the lowest
error on the verification set has reduced, if it has not then the training ends for this iteration. When updating the weights we use the same protocol as Ref.~\citenum{MLCI}  by employing an initial learning rate of 0.1 that drops to 0.01 after the second iteration of MLCI or
if we are using weights from a previous calculation.

MLCI begins by adding all single and double substitutions to the Hartree-Fock single determinant wavefunction while the weights for the ANN are initialized to random values on $[-0.1,0.1]$ if there are no other data provided. We briefly describe the steps in the MLCI algorithm used in this work below:

\begin{itemize}
\item{The Hamiltonian matrix is constructed in the current configuration space then diagonalized to give the wavefunction.}
\item{Configurations with $|c_{i}|<c\subscript{min}$ that are newly added are removed from the configuration space but retained in a reject set. All of the current configuration space is considered for removal every ten iterations.}
\item{The ANN is then trained to predict the transformed coefficients $|\tilde{c}_{i}|$ by combining the reject set with the wavefunction and dividing these data randomly and equally between training and verification sets.}
\item{All symmetry-allowed single and double substitutions are generated from the current configuration space with their ANN prediction calculated on the fly and duplicates efficiently removed.}
\item{Those $L$ configurations, where $L$ is size of the current wavefunction, with the largest predicted $|\tilde{c}_{i}|$, are used to enlarge the current configuration space.}
\item{The procedure is repeated until the energy satisfies the convergence criterion of Ref.~\citenum{GreerMcciSpectra} which is tested on every iteration after the first six.}
\end{itemize}

The transformed coefficients are used to increase the contrast between important and unimportant configurations when training, by ensuring that coefficients greater than the cutoff for removing configurations (c\subscript{min}) become greater than 0.6 as the logistic functions used in the ANN give a value of 0.5 if all weights are zero. In contrast those coefficients less than c\subscript{min} are set to zero. The coefficients are normalized so that $\sum_{i} |c_{i}|^{2}=1$ which means that $|c_{i}| \leq 1$. The transformation for those greater than or equal to c\subscript{min} is given by\cite{MLCI} 
\begin{equation}
|\tilde{c}_{i}| = \frac{0.4|c_{i}|+0.6-c\subscript{min}}{1-c\subscript{min}}.
\end{equation} 
so that coefficients on [c\subscript{min},1] are mapped to [0.6,1]. We use Molpro\cite{MOLPRO_2015} for the calculation of Hartree-Fock molecular orbitals together with the one-electron and two-electron integrals.

\subsection{Artificial neural network as a hash function}

The original MLCI algorithm\cite{MLCI} used the procedure of Ref.~\citenum{MCCInatorb} to remove duplicates when generating all single and double substitutions from the current configuration space. This entailed storing all of them then sorting the list using the quicksort algorithm. This was followed by one pass through the sorted list to delete replicas. If the size of the singles and doubles space is $N_{sd}$ then this tends to scale as $O(N_{sd}\log N_{sd})$. Hence it is efficient but by having to store all of the generated singles and doubles then the scalability of the method is restricted to a maximum size for the singles and doubles space as this has to be stored in memory. 

In this Article we remove this impediment by storing only  the $L$ configurations that are currently predicted to be the most important. This is achieved by generating single and double substitutions on the fly and applying the ANN. If the predicted value is less than the lowest predicted value of the current $L$ most important then the next configuration is considered. Otherwise the configuration is checked to see if it is a duplicate. As we have already calculated it, we use the output of the ANN for this configuration as a hash function to check for duplicates in an efficient way.  The hash value is given as  $\lfloor \text{Output}2L \rfloor+1$ where $2L$ was chosen to keep the size of the array similar to the size of the wavefunction while reducing collisions, i.e., different configurations with the same hash value. For example, if the ANN predicts 0.8643 as the importance of the configuration and there are $1000$ configurations in the wavefunction
then we only have to check that the new configuration is not a duplicate of the existing configurations, if any, stored at 1729 in the hash table for the $L$ most important configurations and for the current wavefunction. If it is not a duplicate then the least important of the $L$ stored configurations is deleted, the new configuration is added to the ordered list and the hash table is updated. The next single and double substitution is then generated and the process is continued until the space has been exhausted.

While using the ANN also as a hash function to delete duplicates on the fly removes the memory bottleneck of storing all of the singles and doubles, it could be slower than the previous approach as now the ANN is applied to all of the duplicates as well as the unique singles and doubles space. However the hypothesis is that because applying the ANN is fast then this should not impose too much of a burden and we shall see in the Results section that using the ANN as a hash function can actually enable MLCI to run faster than when using the quicksort approach to delete duplicates.

\subsection{Configuration state functions}

The MLCI procedure is also built upon to work with configuration state functions (CSFs) so that the computed wavefunctions will be pure states for a given total spin. This uses the framework of the MCCI program\cite{MCCIGreer95,mcciGreer98,mccicodeGreer}

to calculate the now more involved matrix elements for the Hamiltonian matrix $\bm{H}$ and, as the CSFs are not orthonormal, their overlap matrix $\bm{S}$. Then the matrix equation $\bm{H}\vec{c}=E\bm{S}\vec{c}$ is solved to give the energy and coefficients.

For a chosen total spin, the $N$ spin orbitals can define an $N$ electron CSF, but the same $N$ spatial orbitals with spins swapped around may correspond to another CSF that is linearly dependent on the first. To prevent this the genealogical procedure of MCCI\cite{mccicodeGreer} is used to put the list of $N$ orbitals into a common form by randomly swapping spins so that the corresponding CSF is either a linearly independent spin eigenfunction or has the same spin orbitals as another CSF so can later be recognized as a duplicate. In the resulting wavefunction, the CSFs are made to be approximately orthonormal using the method of  Ref.~\citenum{GreerMcciSpectra} which scales the coefficients of the wavefunction using  $c'_{i}=c_{i}\sqrt{S_{ii}}/\sqrt{\sum_{j} c_{j}S_{jj} c_{j}}$ and means that $\sum_{i} |c'_{i}|^{2}=1$.

As the occupied spin orbitals now uniquely define a linearly independent CSF then it should be possible for the ANN to use these occupations as an input to learn to predict the importance of CSFs and the Results section demonstrates that this can indeed be successfully accomplished.

\subsection{Transferability}

The idea of transferability is of interest for the efficient application of machine learning, i.e., how well does the trained algorithm transfer to a different but related task. This concept, for example, has been investigated\cite{Welborn18}  in electronic structure when predicting CCSD and MP2 correlation energies for 
transfers that range from between geometries to among molecules and elements. In this work we consider transferability between geometries and take a broader view, as in MLCI the ANN is embedded inside a selected CI scheme, that encompasses other data not just the weights of the ANN. We investigate whether transferring data from the 
previous geometry as we stretch a bond can offer any benefits such as accelerating an MLCI calculation or improving its accuracy. Three transfer protocols are considered which are compared with standard MLCI. In the first we transfer the weights of the ANN, the reject set and the wavefunction as a starting point for the next MLCI calculation. The second protocol only transfers the weights of the ANN while the third only transfers the wavefunction ($\Psi$). We emphasize that the ANN keeps learning in all of these protocols, but its initial weights are the trained weights of a previous geometry when the ANN weights are transferred and the initial training rate is lowered to 0.01 in this case.

\section{Results}

We first compare MLCI using the ANN as a hash function with the previous approach that sorted all of the singles and doubles to remove duplicates. For this initial comparison we use Slater determinants, $30$ hidden nodes and fix the seed for the random number generator to allow a fair comparison between the ways to remove duplicates. We consider the ground state of N\subscript{2} in the cc-pVDZ basis set with two frozen orbitals and use a cutoff of c\subscript{min}$=5\times10^{-4}$. 

\begin{figure}[h!]\centering
\includegraphics[width=.6\textwidth]{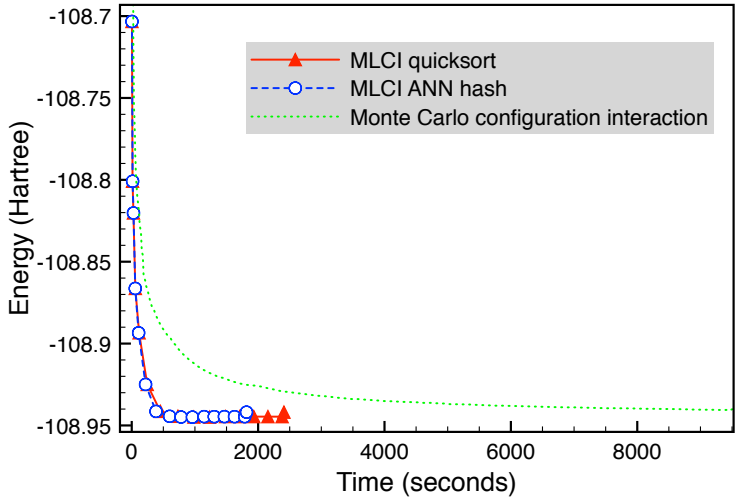}
\caption{Energy (Hartree) against time (seconds) for serial calculations on N\subscript{2} at a bond length of $2.2225$ \AA~ in the cc-pVDZ basis with two frozen orbital using MLCI with quicksort to remove duplicate configurations, MLCI with the ANN as a hash function to remove duplicate configurations, or MCCI. Here all methods employ Slater determinants and a cutoff of c\subscript{min}$=5\times10^{-4}$.}\label{fig:SinglePointN2}
\end{figure}

For a stretched bond length of $2.2225$ \AA~we see in Fig.~\ref{fig:SinglePointN2} that the more scalable new approach to duplicates in MLCI gives the same results as before, but in less time. The final energies were identical yet using the ANN prediction as a hash function required 1816 seconds compared with 2409 seconds when using quicksort to remove duplicates. Both versions are also significantly faster than stochastically selecting configurations using Monte Carlo configuration interaction (MCCI)\cite{MCCIGreer95,mcciGreer98,mccicodeGreer} with Slater determinants when running on a single processor: MCCI used 9504 seconds to give an energy that is slightly higher by 1.1 kcal/mol and required 13502 determinants compared with 12456 in MLCI.

\subsection{Nitrogen molecule potential curve}
We now continue with the new version of MLCI that employs the ANN as a hash function, and use CSFs, rather than Slater determinants, to guarantee that the wavefunction is a pure spin state. We apply this approach to potential energy curves beginning 
with the ground-state singlet of N\subscript{2} in the cc-pVDZ basis set with two frozen orbitals. The FCI results used for comparison were calculated for this system in Refs.~\citenum{LarsenN2FCI2000,chanN2FCI2002,GwaltneyN2FCI2002}. The potential curve of N\subscript{2} is known to be a challenging problem as, for example, coupled cluster methods that use a single reference can perform poorly when the bond is stretched.\cite{RevModPhysBartlett07}

\begin{figure}[h!]\centering
\includegraphics[width=.6\textwidth]{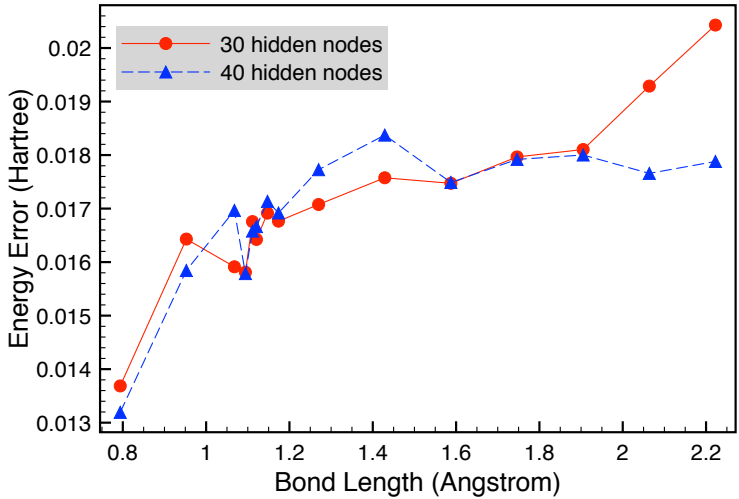}
\caption{Difference between the FCI and MLCI energy (Hartree) for N\subscript{2} with a stretched geometry of $2.2225$ \AA~ in the cc-pVDZ basis set with two frozen orbitals where for MLCI the number of hidden nodes is varied, CSFs are used and c\subscript{min}$=5\times10^{-4}$.}\label{fig:N2curvedifferences}
\end{figure}

The number of hidden nodes is first considered and we run calculations with a cutoff of c\subscript{min}$=5\times10^{-4}$. In Fig.~\ref{fig:N2curvedifferences} we plot the energy error for the MLCI result when compared with the FCI data
for the $15$ bond lengths. We see that increasing the number of hidden nodes from $30$ to $40$ noticeably reduces the error at the most stretched geometries,
but slightly increases the error at some intermediate bond lengths. Both MLCI potential curves used around 4900 CSFs on average while the FCI wavefunctions required $\sim5\times 10^{8}$ determinants. Hence
it is not expected that the compact wavefunctions from MLCI will give energies that are extremely close to FCI, but rather that errors will be sufficiently balanced and so relative energies will be much more accurate. As a potential curve can be shifted by a constant energy, we can aim for MLCI potential energy curves that are essentially FCI quality. 

\begin{figure}[h!]\centering
\includegraphics[width=.6\textwidth]{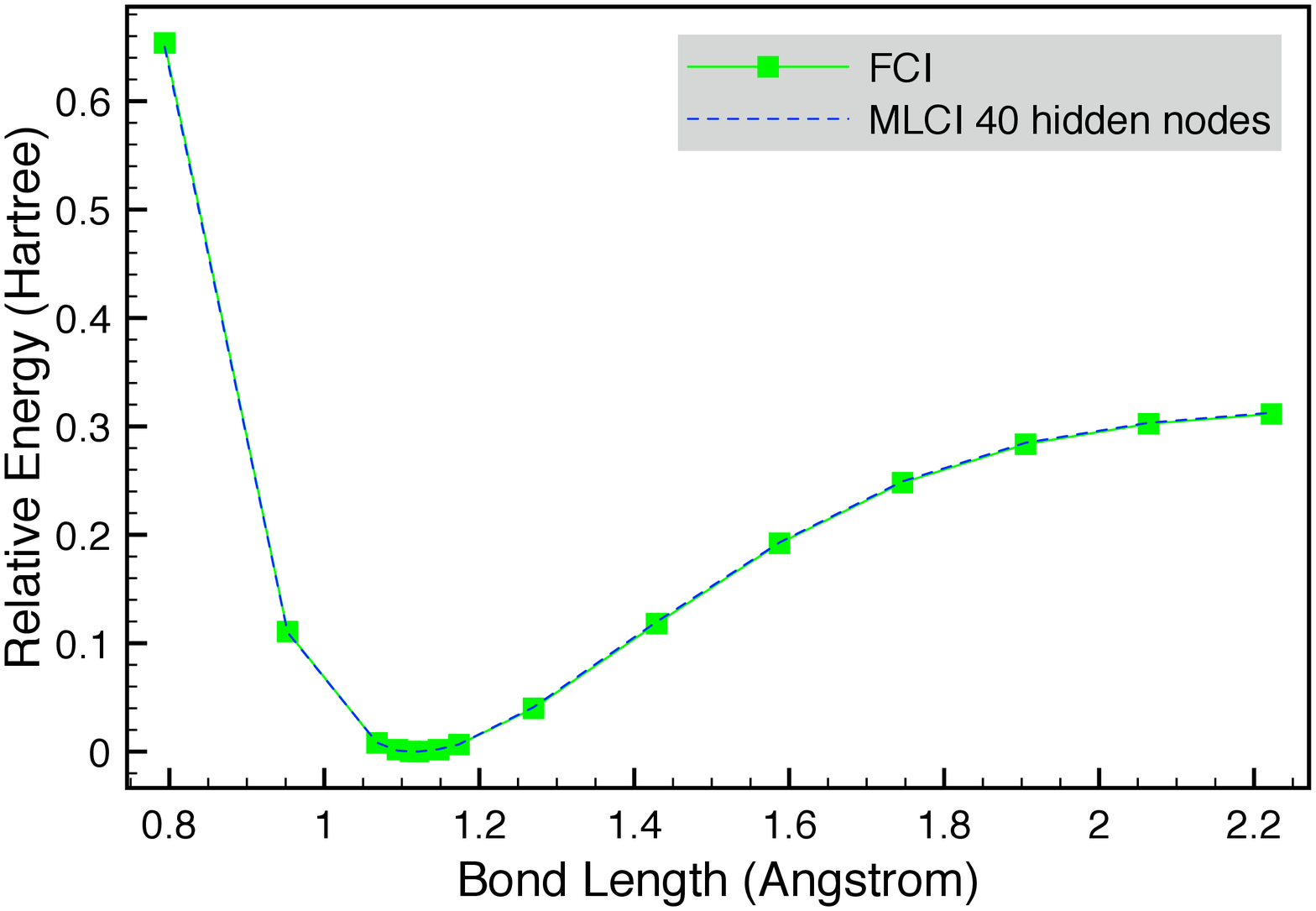}
\caption{FCI and MLCI energies relative to their minimum values against bond length for N\subscript{2} in the cc-pVDZ basis set with two frozen orbitals when using MLCI with 40 hidden nodes, CSFs and c\subscript{min}$=5\times10^{-4}$.}\label{fig:N2curveMLCI_shift}
\end{figure}

 If we shift the potential curves so that they all have their minimum at zero, we see in Fig.~\ref{fig:N2curveMLCI_shift} that with c\subscript{min}$=5\times10^{-4}$ and 40 hidden nodes the MLCI potential curve is practically indistinguishable from the FCI result
on the scale of the graph. To indicate the multireference character of the MLCI wavefunction for the given orbitals we use 

\begin{equation}
MR = \sum_{i}  |c_i|^{2} - |c_i|^{4}
\end{equation}
from Refs.~\citenum{MCCImetaldimers} and \citenum{MRinQC}.  $MR$ is zero for a single determinant and approaches one as the number and importance of the configurations increases. Here the $c_i$ are the coefficients of the configuration interaction wavefunction normalized so that $\sum_{i} |c_i|^{2}=1$. This approach is approximate when using non-orthonormal CSFs. Although the CSFs are made approximately orthonormal in MLCI using the procedure of Ref.~\citenum{GreerMcciSpectra}, the time cost is not too important for individual $MR$ calculations so we make the CSFs exactly orthonormal for use with this indicator. To do this we use the approach
of Ref.~\citenum{mcciopenshell}, which means that the new coefficients are $\vec{c}_{\text{new}}=\sqrt{\bm{S}} \vec{c}$ where $\bm{S}$ is the overlap matrix for the original non-orthonormal CSFs. We find that for the shortest bond length considered then
$MR=0.121$ for the orthonormal CSFs suggesting that single-reference approaches are likely to work sufficiently well while this increases to $0.908$ at the longest bond length indicating that the problem, with Hartree-Fock orbitals, is now very strongly multireference.

To enable a quantitative comparison of the accuracy of approximations to the FCI curve and take into account that the potential curves can be shifted by a constant we use the 
non-parallelity error\cite{doi:10.1063/1.469812} (NPE) and the standard deviation of the difference in energies\cite{MCCInatorb} $\sigma_{\Delta E}$. 

If we write $\Delta E_{i}=E_{i,approx}-E_{i,FCI}$ then
\begin{equation}
NPE=\max_{i} |\Delta E_{i} | - \min_{i} |\Delta E_{i}|
\end{equation}
where $i$ ranges over all points in the curve. Hence if the approximate curve when shifted by a constant is equal to the FCI curve then $NPE=0$. The standard deviation of the difference in energies is also equal to zero when there is a constant difference between the curves and furthermore depends on all points in the curve not just the maximum and minimum $\Delta E_{i}$. It is defined as
\begin{equation}
\sigma_{\Delta E}=\sqrt{\frac{1}{d}\sum_{i=1}^{d}(\Delta E_{i}-\overline{\Delta E})^{2}}
\end{equation}
where $d$ is the number of points in the curve and $\overline{\Delta E}$ is the mean of $\Delta E$. 

In Table \ref{tbl:N2} we show the effect of increasing the number of hidden nodes then of transferring all of the ANN, reject list and the wavefunction as the starting point for the next calculation at increasing bond length. 
We also look at transferring just the ANN or only the wavefunction. We see that the errors for the curves when quantified by $\sigma_{\Delta E}$ are generally less than $1$ kcal/mol. On increasing the number of hidden nodes from 30 to 40 the NPE drops by around $1$ kcal/mol with only a modest increase in calculation time. The accuracy of the curve is further improved by transferring all of the wavefunction ($\Psi$), reject list and ANN. However this increases the calculation time by around 1 hour.  
Interestingly if only the ANN is transferred then, although the calculation time is essentially the same as without transferring, the error in the curve is slightly higher than that from the other approaches at a still reasonable 1.12 kcal/mol for $\sigma_{\Delta E}$. The highest
accuracy and longest calculation time is from transferring only the wavefunction and this give $\sigma_{\Delta E}=0.56$ kcal/mol but took almost 1.5 hours longer than the standard $n_{h}=40$ calculation. This increase in accuracy and time appears due to a greater number of important configurations being found as on average there are around 1000 more CSFs than in the other approaches. 

\begin{table}[h!]
\centering
\caption{Results for MLCI with  c\subscript{min}$=5\times10^{-4}$ when varying the number of hidden nodes ($n_{h}$) and the transfer protocol when applied to the potential curve of N\subscript{2} in the cc-pVDZ basis set with two frozen orbitals.}
\begin{tabular}{@{\extracolsep{\fill}}lcccc}
\hline
 	 & NPE (kcal/mol)  & $\sigma_{\Delta E}$ (kcal/mol) & Mean CSFs & Time (Hours)   \\
\hline
$n_{h}=30$  &     4.23 & 0.94 & 4904 & 2.90 \\
$n_{h}=40$ &   3.25 & 0.78 & 4856 & 3.12 \\
$n_{h}=40$ Transfer all & 3.09 & 0.71 & 5094 & 4.15 \\
$n_{h}=40$ Transfer ANN  & 4.86 & 1.12 & 4748 & 3.20 \\
$n_{h}=40$ Transfer $\Psi$ & 2.00 & 0.56 & 5972 & 4.57 \\
\hline
\label{tbl:N2}
\end{tabular}
\end{table}

We note that in previous work\cite{MCCInatorb} the stochastic approach of MCCI required 7.55 hours with c\subscript{min}$=5\times10^{-4}$  on the same
hardware when using the Hartree-Fock molecular orbitals to reach an accuracy of 1.06 kcal/mol for $\sigma_{\Delta E}$. This was on 12 processors so equates to 90.6 processor hours. The results show that for the nitrogen dimer, MLCI can produce a more accurate curve in less time compared with randomly selecting configurations using MCCI. It appears that the ANN does not transfer particularly well as a starting point for the geometries of N\subscript{2} considered here. This suggests that the most accurate approach is to use the wavefunction from the previous geometry as 
a starting point and train a completely fresh ANN each time. We continue with 40 hidden nodes and next investigate the performance of MLCI together with transfer protocols on the potential curve for the double hydrogen dissociation of the water molecule.

\subsection{Water potential curve}

\begin{figure}[h!]\centering
\includegraphics[width=.6\textwidth]{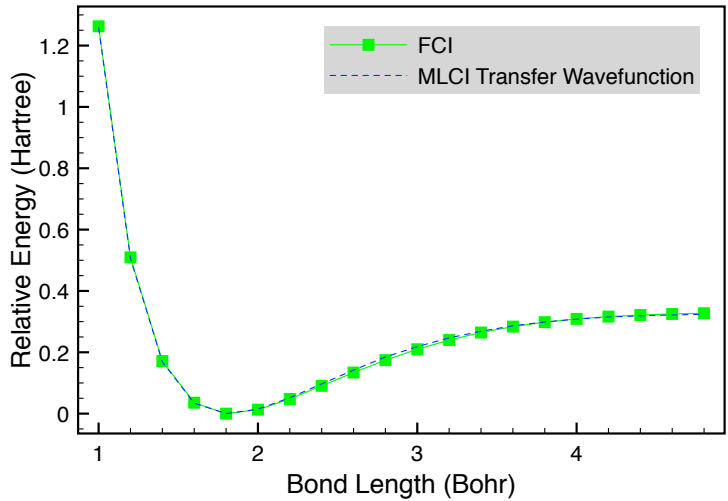}
\caption{FCI and MLCI energies relative to their minimum values for H\subscript{2}O as both bonds are varied in the cc-pVDZ basis set with one frozen orbital when using MLCI with transfer of the wavefunction, 40 hidden nodes, CSFs and c\subscript{min}$=10^{-3}$.}\label{fig:H2OcurveMLCI_shift}
\end{figure}

For the H\subscript{2}O potential energy curves of the ground-state singlet the bond angle is fixed at 104.5 degrees as we simultaneously vary both bond lengths when using the cc-pVDZ basis with one frozen orbital. For comparison we use the FCI results
from Ref.~\citenum{MCCInatorb}. When initially using c\subscript{min}$=10^{-3}$ and transferring $\Psi$ we see in Fig.~\ref{fig:H2OcurveMLCI_shift} that the agreement with FCI is very good on the scale of the graph when both curves are shifted so their minimum is
at zero. However there is a slight difference that can be noticed at intermediate bond lengths.

\begin{figure}[h!]\centering
\includegraphics[width=.6\textwidth]{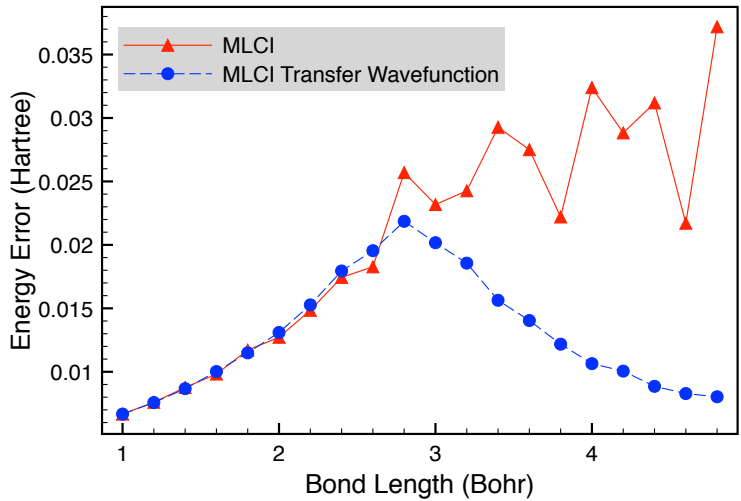}
\caption{Difference between the FCI and MLCI energies with and without transference of the wavefunction for H\subscript{2}O when both bonds are varied in the cc-pVDZ basis set with one frozen orbital when using MLCI with 40 hidden nodes, CSFs and c\subscript{min}$=10^{-3}$.}\label{fig:H2Odifferences}
\end{figure}

The lower accuracy at intermediate bond lengths is confirmed in the plot of differences with the FCI result (Fig.~\ref{fig:H2Odifferences}) where we see that the largest difference when transferring the
wavefunction is at 2.8 Bohr. Fig.~\ref{fig:H2Odifferences} also reveals that if we do not transfer the wavefunction then the errors generally continue rising as the bonds lengthen.

We first consider the multireference character using $MR$ when we do not transfer the MLCI wavefunction. At $1$ Bohr this is $0.065$, by $2.8$ Bohr it has risen to $0.210$  then at the longest bond length considered ($4.8$ Bohr) it is 0.784. We compare this to the results when the wavefunction is transferred, at $1$ Bohr there is nothing to transfer so the value remains the same, at $2.8$ Bohr it is $0.228$ then at $4.8$ Bohr it is 0.808. Hence at the shortest bond length then the problem is certainly single-reference, the $MR$ value then suggests that by 2.8 Bohr there may now be a small amount of multireference character, while the system with Hartree-Fock orbitals has become strongly multireference by $4.8$ Bohr. Although the energy differences are apparent at long bond lengths between the two approaches, both give approximately the same multireference character. That MLCI when transferring the wavefunction performs better after 2.6 Bohr could be due to the increase in multireference character so the algorithm has a better chance to find a sufficient number of the important configurations if it is given the previous set of configurations as a starting point to build and learn from rather than beginning anew from a single determinant.

In table \ref{tbl:H2OlargerCutoff} we compare the accuracy of the curves from different transfer protocols when using c\subscript{min}$=10^{-3}$  and see that, as for N\subscript{2}, transferring the wavefunction substantially improves the results while the least accurate curve
is due to only transferring the ANN. The improvement is more apparent as the NPE values are higher than for N\subscript{2} which we attribute to the larger cutoff used. The most accurate result used the most configurations
on average and consequently required the longest time. We note that previous work\cite{MCCInatorb} with MCCI produced a curve with an error of $\sigma_{\Delta E}=0.92$ kcal/mol in 0.52 hours which equates to 6.2 processor hours. Hence for the water molecule at this cutoff then MLCI is less
accurate which may be due to the multireference character being smaller than N\subscript{2} with regards to the shortest and longest bonds considered, and the FCI space ($\sim 2\times 10^{7}$ determinants) being around an order of magnitude lower. Hence fewer configurations are required for the MLCI wavefunction thus providing less opportunity for learning. To investigate this latter conjecture we next consider lowering the cutoff.  

\begin{table}[h!]
\centering
\caption{Results for MLCI with a cutoff of  c\subscript{min}$=10^{-3}$ and 40 hidden nodes when the transfer protocol is varied when applied to the potential curve for the double hydrogen dissociation of H\subscript{2}O in the cc-pVDZ basis set with one frozen orbital.}
\begin{tabular}{@{\extracolsep{\fill}}lcccc}
\hline
 	 & NPE (kcal/mol)  & $\sigma_{\Delta E}$ (kcal/mol) & Mean CSFs & Time (Hours)   \\
\hline
MLCI  &     19.15 & 5.57 & 1032 & 0.61 \\
MLCI Transfer all & 24.26     & 7.74 & 937 & 0.63 \\
MLCI Transfer ANN  &     26.55 & 8.39 & 930 & 0.57 \\
MLCI Transfer $\Psi$ &     9.53 & 2.87 & 1212 & 0.74  \\
\hline
\label{tbl:H2OlargerCutoff}
\end{tabular}
\end{table}

We see in table \ref{tbl:H2OlowerCutoff} that at the lower cutoff of $5\times10^{-4}$ then the errors are much lower with $\sigma_{\Delta E}$ less than $2$ kcal/mol for all the MLCI approaches and the mean number of CSFs is larger but still less than N\subscript{2}. Now transferring all the data, not just the wavefunction, gives the most accurate curve but the differences between the approaches are small. Transferring the wavefunction required the most CSFs on average and although every individual energy is lower than just transferring the ANN the $\sigma_{\Delta E}$ error is higher. This serves to remind us that it is the consistency in describing multiple points on the potential curve that is important not just getting all points to be lower in energy. Although the results for this system are now sufficiently accurate for all of the MLCI approaches, the maximum $\sigma_{\Delta E}$ of 1.75 kcal/mol is still higher than the c\subscript{min}$=10^{-3}$ MCCI value\cite{MCCInatorb} of 0.92 kcal/mol. We note that we can achieve a 
more accurate potential curve if we further lower the cutoff to c\subscript{min}=$2\times10^{-4}$ where now MLCI when only transferring the wavefunction gives an error of  $\sigma_{\Delta E}=0.70$ kcal/mol but takes longer (13.3 hours) than the c\subscript{min}$=10^{-3}$ MCCI result. In this case more CSFs were used on average (9744) than for the N\subscript{2} results, which fits in with the idea that there needs to a sufficient number of configurations in the wavefunction for the training of the ANN of to work well.
Although the MLCI c\subscript{min}$=5\times10^{-4}$ results were not quite as accurate as the MCCI c\subscript{min}$=10^{-3}$ values, they required less processor hours but the performance gain was not as substantial as for the nitrogen dimer. We attribute this to the lower multireference character of this system 
and fewer configurations being needed for a given cut-off so the MCCI calculations did not need much time and there is less scope for improvement.

\begin{table}[h!]
\centering
\caption{Results for MLCI with a cutoff of  c\subscript{min}$=5\times10^{-4}$ and 40 hidden nodes when the transfer protocol is varied when applied to the potential energy curve for the double hydrogen dissociation of H\subscript{2}O in the cc-pVDZ basis set with one frozen orbital.}
\begin{tabular}{@{\extracolsep{\fill}}lcccc}
\hline
 	 & NPE (kcal/mol)  & $\sigma_{\Delta E}$ (kcal/mol) & Mean CSFs & Time (Hours)   \\
\hline
MLCI  &     6.68 & 1.54 & 2256 & 1.59 \\
MLCI Transfer all & 5.74     & 1.39 & 2430 & 2.15 \\
MLCI Transfer ANN  & 7.27    & 1.67 & 2226 & 1.60 \\
MLCI Transfer $\Psi$ &    6.05 & 1.75 & 2892 & 2.28  \\
\hline
\label{tbl:H2OlowerCutoff}
\end{tabular}
\end{table}

\subsection{Carbon monoxide potential curve}

\begin{figure}[h!]\centering
\includegraphics[width=.6\textwidth]{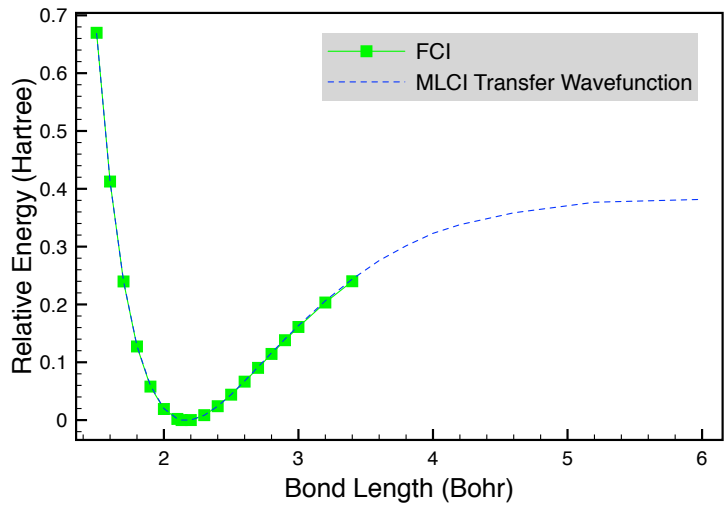}
\caption{FCI and MLCI energies relative to their minimum values against bond length for CO in the cc-pVDZ basis set with two frozen orbitals when using MLCI with transfer of the wavefunction, 40 hidden nodes, CSFs and c\subscript{min}$=5\times10^{-4}$.}\label{fig:COcurveMLCI_shift}
\end{figure}

Finally we consider the potential curve for carbon monoxide when using the cc-pVDZ basis with two frozen orbitals. We see in Fig.~\ref{fig:COcurveMLCI_shift} that, when using c\subscript{min}$=5\times10^{-4}$ and transferring the wavefunction, the MLCI potential curve
is essentially FCI quality for the points where FCI values are available. We note that there are now $\sim 10^{9}$ determinants in the FCI space which makes it the largest considered in this Article, and at a geometry of $6$ Bohr we could not run a Molpro\cite{MOLPRO_2015} FCI calculation
to completion as it required more than 1TB of disk space which we partly attribute to the strong multireference character. Hence we use the FCI values from Ref.~\citenum{MCCInatorb} which are available to a maximum bond length of 3.4 Bohr.    

\begin{figure}[h!]\centering
\includegraphics[width=.6\textwidth]{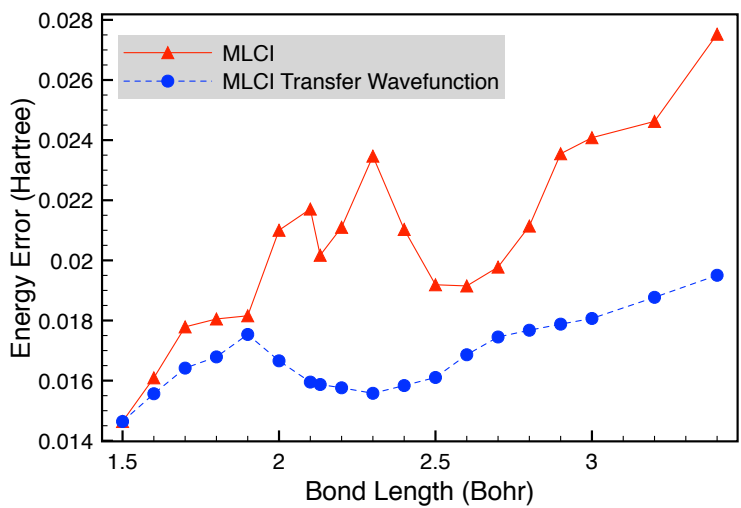}
\caption{Difference between the FCI and MLCI energies against bond length with and without transference of the wavefunction for CO in the cc-pVDZ basis set with two frozen orbitals when using MLCI with 40 hidden nodes, CSFs and c\subscript{min}$=5\times10^{-4}$.}\label{fig:COdifferences}
\end{figure}

In Fig.~\ref{fig:COdifferences} we see that transferring the wavefunction in MLCI gives lower and more consistent errors in comparison to standard MLCI. This improvement in performance appears better
than that for the potential curve of water when the geometries are not too stretched (Fig.~\ref{fig:H2Odifferences}).  

Table \ref{tbl:CO} shows that, as with N\subscript{2}, transferring just the wavefunction gives the most accurate curve with a $\sigma_{\Delta E}$ error that is less than half the value for the other approaches. There is a noticeable time cost in calculations involving transferring the wavefunction which require around 19 processor hours. Now transferring just the ANN is the second most accurate approach, but the errors for transferring all or just the ANN are both reasonably similar to standard MLCI. We note that errors when transferring all data from each point to the next are slightly higher than standard MLCI despite using almost 1000 more CSFs on average. This is due to
the error calculation only being possible for the first 17 points where we have FCI values. For example the MLCI results when transferring all data are actually 0.02 Hartree lower at $R=6$ than the value from standard MLCI.  

We also compare these data with previous results\cite{MCCInatorb} using MCCI. There the $\sigma_{\Delta E}$ error was 0.89 kcal/mol when using Hartree-Fock orbitals, and the time for the entire curve was 20.72 hours which equates to 248.6 processor hours. Hence by transferring only the
wavefunction in MLCI we can achieve slightly higher accuracy with slightly less wall time and with a very significant reduction in the number of processor hours.

\begin{table}[h!]
\centering
\caption{Results for MLCI with a cutoff of  c\subscript{min}$=5\times10^{-4}$ and 40 hidden nodes when the transfer protocol is varied when applied to the potential curve of CO in the cc-pVDZ basis set with two frozen orbitals. Errors are
with regards to the available 17 FCI values with bond lengths less than or equal to $3.4$ Bohr, while CSFs and timings are for the entire curve.}
\begin{tabular}{@{\extracolsep{\fill}}lcccc}
\hline
 	 & NPE (kcal/mol)  & $\sigma_{\Delta E}$ (kcal/mol) & Mean CSFs & Time (Hours)   \\
\hline
MLCI  &     8.08 & 1.91 & 5586 & 12.17 \\
MLCI Transfer all & 8.54   & 2.19 & 6438  & 18.55 \\
MLCI Transfer ANN  &     7.25 & 1.72 & 5880 & 12.31 \\
MLCI Transfer $\Psi$ &     3.05 & 0.76 & 7496 & 19.84 \\
\hline
\label{tbl:CO}
\end{tabular}
\end{table}

As for the previous two systems we again make the CSFs orthonormal to calculate $MR$ values to indicate the multireference character for MLCI when transferring the wavefunction. We find $MR=0.116$ at 1.5 Bohr and 0.192 at 2.1316 Bohr suggesting that it would not yet be classified as multireference at these geometries but there may be a small amount of multireference character at the equilibrium geometry. However by 3.4 Bohr this has risen to 0.736 indicating that it is now a multireference problem and $MR$ reaches the highest value in this Article of $0.931$ at 6 Bohr demonstrating that the problem is very strongly multireference by this point.

\section{Summary} 

In this Article, machine learning configuration interaction (MLCI) was further developed so that the singles and doubles space did not need to be stored when efficiently removing duplicates. This removed a restriction on the scalability of the method and was achieved by 
using the artificial neural network (ANN) prediction as a hash function to allow duplicates to be efficiently found when configurations are generated on the fly. This enhancement of MLCI to make it more scalable
was demonstrated to give the same result as the original MLCI in less time for a single-point energy of the nitrogen molecule when using Slater determinants, and was substantially faster than stochastically selecting configurations using 
Monte Carlo configuration interaction (MCCI) when run in serial.  

MLCI was then applied to the problem of calculating {\it ab initio} potential curves. Here configuration state functions were used in the procedure to ensure that the wavefunction was a pure spin state and the transferability of data 
from the previous geometry as a starting point to improve the next calculation was investigated. 

For the nitrogen molecule in a cc-pVDZ basis we found that the MLCI potential curve was practically indistinguishable to the full configuration interaction (FCI) result when energies were plotted relative to their
minimum value. This accuracy was quantified and supported by using the non-parallelity error and $\sigma_{\Delta E}$ which were as low as 2.00 and 0.56 kcal/mol respectively when transferring the wavefunction. This was more accurate than previous results using MCCI to randomly
select configurations\cite{MCCInatorb}, used less wall time despite MLCI being run in serial while MCCI was run on 12 processors, and substantially lowered the number of processor hours.  Transferring the wavefunction was more accurate than transferring all
of the data from the previous geometry, while only transferring the ANN gave the largest error although not by much and this was still a respectable 1.12 kcal/mol for $\sigma_{\Delta E}$.

The water molecule was then considered and the potential curve from MLCI when using a larger
cutoff of $10^{-3}$ could not be distinguished from FCI on the scale of the graph except for a region around intermediate bond lengths. Transferring the wavefunction
gave a noticeable reduction in errors but they were higher than using MCCI. The errors were lowered on decreasing the cutoff to $5\times10^{-4}$ yet the curves remained slightly less accurate
than when using MCCI with a cutoff of $10^{-3}$, although less processor hours were needed for MLCI. The cutoff had to be lowered to $2\times10^{-4}$ to surpass the accuracy of the MCCI $10^{-3}$ result and now MLCI required more processor hours. It was suggested
that the smaller configuration space and lower multireference character of this system meant that there was less opportunity for the ANN to learn and not so much room for improvement on the already relatively fast MCCI calculations.     

Finally we looked at carbon monoxide when using a cc-pVDZ basis and found that when transferring the wavefunction then MLCI gave essentially a FCI quality potential curve for the points where FCI data were available. 
This MLCI curve was of slightly higher accuracy than previous results using random selection of configurations with MCCI and used slightly less wall time. Furthermore the number of processor hours
used was significantly lowered by MLCI from 248.6 to 19.8. Again transferring solely the wavefunction gave the best results while now only transferring the ANN was the second most accurate protocol, but there was not much difference in accuracy between this and either transferring all data or nothing.

We have seen that this improved, more scalable machine learning configuration interaction can cope with geometries ranging from single reference to very strongly multireference problems to produce potential curves that are practically FCI quality. MLCI generally performed best when transferring just the wavefunction. This protocol, when running in serial, used less wall time and processor hours but gave a more accurate curve than when stochastically selecting configurations in parallel for N\subscript{2} and CO. This was not the case however for the water molecule, which was attributed to its configuration space and multireference character being lower. It was generally seen that transferring the ANN from a near geometry as a starting point did not confer any advantage in accuracy but nor did it significantly change the calculation time, while transferring just the wavefunction could noticeably improve accuracy but required more time. Hence it appeared that the best approach was to have the ANN start learning afresh but begin with the configurations of a near geometry. As only a single hidden later with 40 nodes was used here then training of the ANN does not represent a significant cost, it is possible that transferring the ANN could offer an improvement in time and accuracy when using deep nets trained on graphical processing units which will be investigated in future work when applying the approach to larger molecules. A parallel MLCI will also be considered where, for example, the configurations that have been predicted to be important are not added all at once for a single diagonalization, but shared between smaller concurrent diagonalizations to provide approximate training data on their importance in the wavefunction.

\begin{acknowledgement}
JPC thanks the EPSRC for support via the platform grant EP/P001459/1.
\end{acknowledgement}

\providecommand{\latin}[1]{#1}
\makeatletter
\providecommand{\doi}
  {\begingroup\let\do\@makeother\dospecials
  \catcode`\{=1 \catcode`\}=2\doi@aux}
\providecommand{\doi@aux}[1]{\endgroup\texttt{#1}}
\makeatother
\providecommand*\mcitethebibliography{\thebibliography}
\csname @ifundefined\endcsname{endmcitethebibliography}
  {\let\endmcitethebibliography\endthebibliography}{}

\end{document}